\documentclass{aa}  

\usepackage{graphicx}
\usepackage{txfonts}
\usepackage[normalem]{ulem}
\usepackage{xcolor}
\usepackage{lscape}
\usepackage{comment}
\usepackage{booktabs}
\usepackage{topcapt}
\usepackage{natbib}
\usepackage{lineno}
\bibpunct{(}{)}{;}{a}{}{,} 
\usepackage{hyperref}

\begin{document}
\nolinenumbers

   \title{Accurate sticking coefficient calculation for carbonaceous dust growth through accretion and desorption in astrophysical environments}
   \titlerunning{Carbonaceous cosmic dust growth through accretion}

   \author{D. Bossion
          \inst{1}$^,$\inst{2}
          \and
          A. Sarangi\inst{3}$^,$\inst{4}
          \and
          S. Aalto\inst{5}
          \and
          C. Esmerian\inst{5}
          \and
          S. R. Hashemi\inst{2}
          \and
          K. K. Knudsen\inst{5}
          \and
          W. Vlemmings\inst{5}
          \and
          G. Nyman\inst{2}
          }

   \institute{Institute of Physics of Rennes, UMR-CNRS 6251, University of Rennes, F-35000 Rennes, France\\
              \email{duncan.bossion@univ-rennes.fr}
              \and
              Department of Chemistry and Molecular Biology, University of Gothenburg, SE-405 30, Gothenburg, Sweden\\
              \email{nyman@chem.gu.se}
              \and
              DARK, Niels Bohr Institute, University of Copenhagen, Jagtvej 155A, 2200 Copenhagen, Denmark\\
              \email{sarangi@nbi.ku.dk}
              \and
              Indian Institute of Astrophysics, 100 Feet Rd, Koramangala, Bengaluru, Karnataka 560034, India
              \and
              Department of Space, Earth and Environment, Chalmers University of Technology, SE-412 96, Gothenburg, Sweden
             }

   \date{Received September 15, 1996; accepted March 16, 1997}

 
  \abstract
   {Cosmic dust is ubiquitous in astrophysical environments, where it significantly influences the chemistry and the spectra. Dust grains are likely to grow through the accretion of atoms and molecules from the gas-phase onto them. Despite their importance, only a few studies compute sticking coefficients for relevant temperatures and species, and their direct impact on grain growth. Overall, the formation of dust and its growth are processes not well understood.}
   { To calculate sticking coefficients, binding energies, and grain growth rates over a wide range of temperatures, for various gas species interacting with carbonaceous dust grains.}
   {We perform molecular dynamics simulations with a reactive force field algorithm to compute accurate sticking coefficients and obtain binding energies. The results are included in an astrophysical model of nucleation regions to study dust growth.}
   {We present, for the first time, sticking coefficients of H, H$_2$, C, O, and CO on amorphous carbon structures for temperatures ranging from 50~K to 2250~K. In addition, we estimate the binding energies of H, C, and O in carbonaceous dust to calculate the thermal desorption rates. Combining accretion and desorption allows us to determine an effective accretion rate and sublimation temperature for carbonaceous dust.}
   {We find that sticking coefficients can differ substantially from what is commonly used in astrophysical models and this gives new insight on carbonaceous dust grain growth via accretion in dust-forming regions.}


   \maketitle

\section{Introduction}

Dust grains in space provide a surface on which atoms and molecules may collide and stick, and also undergo chemical reactions. They are also important for the thermodynamics of the interstellar medium (ISM)~\citep{draine_2011}. The evolution of dust grains in these processes depends on several factors, such as the physical conditions of the environment, the chemical type and morphology of the grains, as well as the abundance of the gas-phase species~\citep{dwek_1980, tielens_1998, dwe11, jones_2011}. When a species collides with a dust grain, its probability of sticking to the surface, and hence to participate to its growth by accretion, is called the sticking coefficient~\citep{devlin_1985, laffon_2021}. This factor significantly impacts the evolution and processing of cosmic dust in astrophysical environments~\citep{andersen_iau2003}, especially in regions where dust grains are thought to nucleate from the gas-phase in so-called nucleation regions - the winds of Asymptotic Giant Branch (AGB) stars and the ejecta of Core-Collapse Supernovae (CCSNe)~\citep{bocchio_aa2014}. Given a lack of theoretical predictions or experimental measurements, the previous literature has mainly treated the sticking coefficient as a parameter. In most cases, a fixed value has been chosen, e.g., 1 or 0.5, regardless of the type of gas particle, of the dust composition, or of the gas conditions \citep{dwe11,hirashita_2012}. Here we emphasize that the fate of cosmic dust largely depends on the sticking coefficients and requires a more accurate knowledge of such quantities. 

Although the chemical and mineral composition of cosmic dust remains uncertain, both empirical constraints and theoretical considerations suggest that, especially in nucleation environments, cosmic dust consists mainly of two different species: carbonaceous and silicate dust~\citep{tielens_fass2022}. In O-rich nucleation environments, silicate dust is found and is composed of various species, but mainly O, Si, Fe, Mg, Ca, and Al~\citep{henning_araa2010,gobrecht_iau2019}, since most C is trapped in CO molecules. In C-rich nucleation environments, carbonaceous dust is formed and is composed mainly of C and H, the O being trapped mostly in CO molecules. In this work, we are interested in the latter dust type. 

Due to the importance of accretion that is, together with coagulation, the main grain growth mechanism, previous works have obtained sticking coefficients for some silicate and carbonaceous grains both theoretically~\citep{devlin_1985} and experimentally~\citep{chaabouni_aa2012}. Concerning carbonaceous dust, calculations were carried out mainly on graphite-like surfaces~\citep{devlin_1985,hollenbach_jcp1970,hollenbach_apj1971,cazaux_aa2011}, which is one possible structure of carbon dust. However, spectroscopic evidence supports that carbonaceous grains are often in the form of amorphous carbon structures~\citep{ehrenfreund_cshlp2010,nashimoto_apjl2020,cherchneff_iau1999} for which very little literature on sticking coefficients exists, and that which does exist only concerns hydrogenated amorphous carbon~\citep{rooij_pccp2010,vonkeudell_2002}, due to its importance in materials science. Many studies also focus on dust in cold environments (below 100~K), where the dust particles in circumstellar ejecta/nucleation environments can be partially coated with amorphous ice, for which more calculations and experiments are available~\citep{buch_apj1991,masuda_asr1997,alhalabi_jcp2004,chaabouni_aa2012,veeraghattam_aj2014}. 

In the present work, we study the sticking process of different atomic and diatomic species on amorphous carbon particles. We compute the sticking coefficients as a function of temperature, ranging from 50~K to 2250~K, setting the dust and gas at the same temperature. In fact, this study provides sticking coefficients, not available elsewhere in the literature, that are fundamental to model environments where carbon dust is present without an icy mantle and is in thermal equilibrium with the gas-phase, i.e., primarily the circumstellar ejecta in which dust grains nucleate. Using these coefficients, we also show how the gas-phase chemistry will impact the growth of these grains. 

In the following section \ref{num_det}, Numerical Details, we give some details on the amorphous carbon structure used and the dynamical calculations performed to determine sticking coefficients and binding energies. Thereafter comes the section \ref{res_disc}, Results and Discussion, where we present the obtained sticking coefficients and an application to the dust growth process in astrophysical environments, which is followed by a discussion of the impact of this work and finally by section \ref{concl}, Conclusions. 

\section{Numerical details}
\label{num_det}
In this section we describe the amorphous carbon grain used in our simulations, how the simulations are performed and how the binding energies are obtained.
\subsection{The amorphous carbon structure}

\begin{figure*}[htbp]
\sidecaption
\includegraphics[width=12cm]{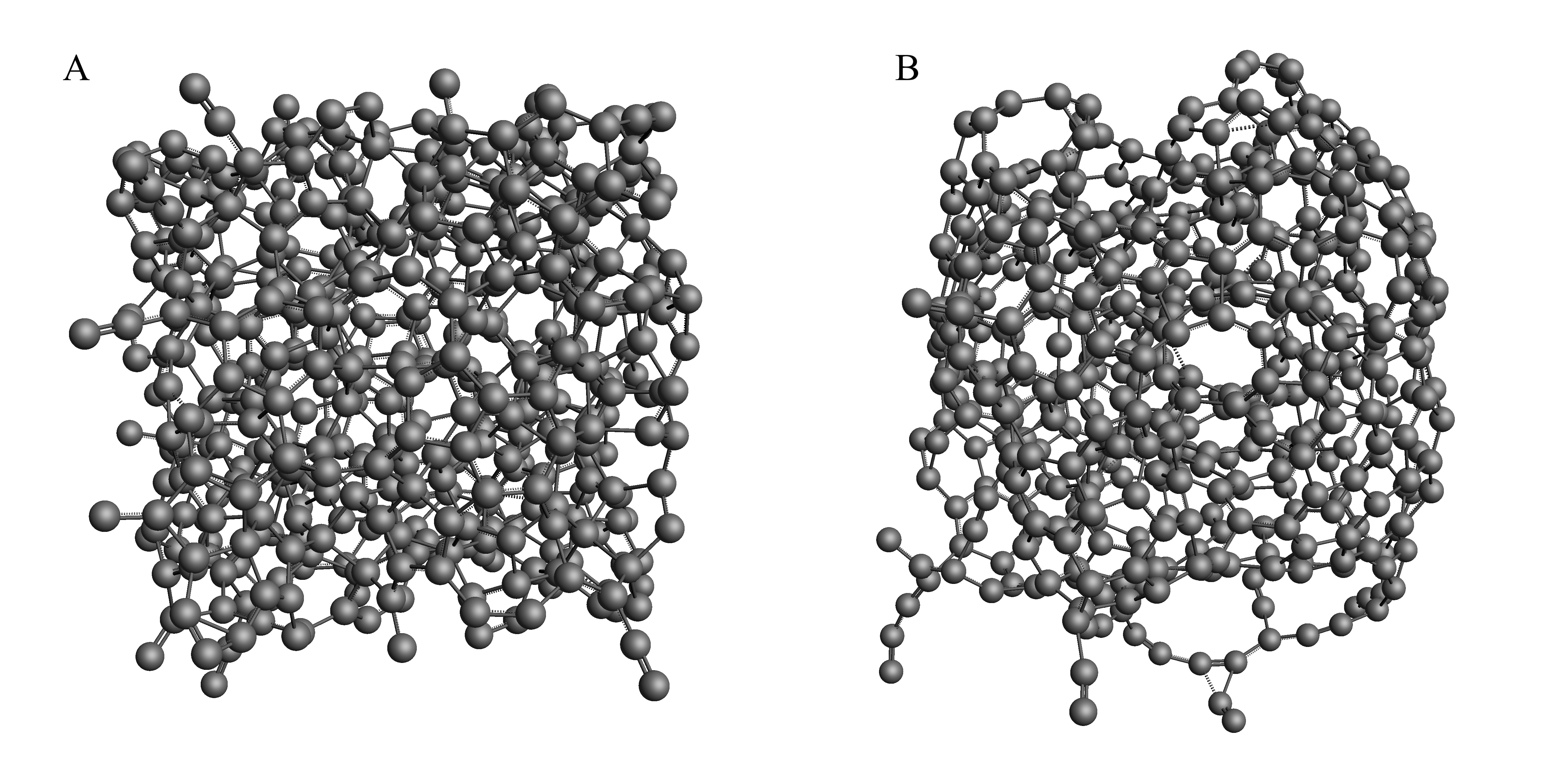} 
\caption{\textbf{Thermalized amorphous carbon structure.} Panel A (left): at 50~K, presenting its original die shape, panel B (right): at 2500~K, presenting many two-dimensional cycles.}
\label{fig:aCtherm}
\end{figure*}

In this work, we focus on carbonaceous cosmic dust. This dust is mainly composed of amorphous carbon, and hence this is the initial structure that we consider to compute the sticking coefficients. The structures used come from a library of tetrahedral amorphous carbon structures available in the literature\citep{deringer_CM2018}. Those structures were obtained by combining molecular dynamics (MD) with a Gaussian approximation potential (GAP) based on machine learning (ML)\citep{deringer_PRB2017}. That ML-GAP procedure provides close-to-density functional theory (DFT) accuracy for a low computational cost. The authors created structures by taking slabs from an initial bulk obtained by GAP. They inserted a perturbation (by inserting a vacuum region) into those slabs. Then by MD, driven by ML-GAP, they heated the surface from 300 K to 1000 K. This was followed by annealing at 1000 K and quenching at 300 K. We use their largest amorphous carbon structures available, consisting of 512 carbon atoms, as we aim to simulate atoms and diatomic molecules hitting a carbon dust structure composed of hundred of atoms. We selected 10 independent structures to obtain sticking coefficients averaged over those structures and obtain results that are as independent of the chosen amorphous carbon structures as possible and widely usable.

In order to use the described structures as carbonaceous dust, we first thermalize them to the temperature we are interested in, using a force field and a Nos\'e-Hoover chain\citep{nose_jcp1984,hoover_pra1985}. The sticking coefficient data are necessary for astrophysicists wherever carbonaceous dust is present. In dense environments in which dust grains form, at low temperature, below several tens of kelvins, an icy mantle forms around dust grains and changes the surface on which the incident particles hit. These mantles likely prevent the accretion of gas-phase atoms onto the solid dust substrate, and therefore we do not consider lower temperatures. Here, we choose to start at 50~K where at least part of the grains should not have an icy mantle \citep{marchione_aesc2019,potapov_prl2020}. Moreover, in most astrophysical models of environments that include dust, it is completely sublimated above 2000~K. Consequently, we focus on providing sticking coefficients in the 50-2250~K temperature range. We present in Fig.~\ref{fig:aCtherm} how resilient to temperature these amorphous carbon structures are by thermalizing one structure at 50~K (panel A) and at 2500~K (panel B). At 50~K the structure is similar to the original tetrahedral amorphous carbon structure, while at 2500~K it looses its die shape and includes mostly two-dimensional cyclic units instead of three-dimensional bonds. This is due to the possibility of higher strength of bonds in cyclic compounds, making these shapes more resistant to high temperatures.

\subsection{Dynamical calculation of sticking coefficients}

We aim at performing high-accuracy simulations of the dynamics of incident particles hitting carbon dust. Due to the high number of atoms present during the simulation (512 plus the incoming atoms), we choose to simulate the collisions using classical molecular dynamics. This approach is also consistent with the way the amorphous carbon structures were obtained. The movement of all the atoms and molecules is hence modeled using Newton's equations of motion, whereas all the forces are obtained based on force fields. To perform these simulations, we use, in particular, the reactive force field method (ReaxFF)\citep{duin_jpca2001} that allows bond formation and bond breaking, as we expect this to occur during atom sticking, together with the functional that was developed for hydrocarbon oxidation\citep{chenoweth_jpca2008} by the authors of this method. They optimized the parameters of this force field against quantum mechanical calculations. The latter were validated to yield an average error of 3.11~kcal.mol$^{-1}$ for enthalpies of formation at 298~K. Our calculations were performed using the ReaxFF module of the Amsterdam Modeling Suite (AMS)\citep{reaxFF_SCM}, which is an optimization of the original ReaxFF code.

The environments in which cosmic dust is found include diffuse clouds (typically with densities up to 10$^{4}$~cm$^{-3}$)\citep{potapov_irpc2021}. We can expect that each collision between an incident particle and the dust surface is independent of any earlier or later incoming particle. For this reason, we do not want the different incident particles to interact with each other during our simulation. Hence, for every simulation on a given amorphous carbon structure, we launch colliding particles every 1250~fs, and limit to only 37 particles in total to avoid the initial pure amorphous carbon structure to be too far from its original structure for all incident particles (it will contain at most 37 non-carbon atoms compared to the 512 carbon atoms in the cluster). To increase the statistics, we run the simulation twice, leading to a total of 74 incident particles hitting each of the 10 amorphous carbon structures that we use, for a given temperature. Thus, each sticking coefficient is obtained based on a total of 740 collisions. It is important to note that this limits the accuracy of our sticking coefficients as we will not be able to obtain sticking coefficients with values lower than $1.10^{-3}$. In addition, each incident particle is initially randomly placed in front of one of the 6 faces of the die-shaped amorphous carbon structure, at least 10~\AA~away from the closest carbon atom. This ensures that we maximize the number of amorphous carbon surfaces with which the incident particles will collide. The collision angle between the incident particles and the amorphous carbon structure is also randomly sampled. 

In order to obtain the sticking coefficients corresponding to an environment at a given temperature $T$, we initially thermalize the amorphous carbon structure. Although the initial velocity of the incident particle in a medium at temperature $T$ is supposed to follow the Maxwell-Boltzmann velocity distribution $f(v)$, it is important to note that colliding particles take a velocity-weighting factor as colliding particles with a high velocity collide more often than those with a low velocity. Hence, the velocity distribution that we consider in this work is
\begin{equation}
    vf(v) = v\sqrt{\frac{2}{\pi}}\left(\frac{m}{k_BT}\right)^{3/2}v^2\exp\left(-\frac{mv^2}{2k_BT}\right).
\end{equation}
The AMS software does not allow to directly sample velocities based on an arbitrary function, but only with Gaussian sampling, given a mean and a variance. The $vf(v)$ being already close to a Gaussian, we fit a Gaussian onto this distribution for each colliding particle and temperature and use the fitting parameters in our simulations. For incident diatomic molecules, we additionally give the individual atoms of the diatom an energy corresponding to the temperature $T$. This affects only its rotational and vibrational degrees of freedom. With this procedure, both the initial amorphous carbon structures and the incident particles are properly thermalized. 

The 512 carbon atom structures being not as large as carbonaceous dust particles (which can be as large as 0.1 $\mu$m in the environments considered in this study), we also thermalize the amorphous carbon structure during the dynamics with a Nos\'e-Hoover chain every 100~fs. This ensures that the energy of the incident particles does not change the temperature of the amorphous carbonaceous dust grain, which is equivalent to having dust particles much larger than the incident particles or having enough time between collisions to thermalize the structure. To ensure the best convergence of the chemisorbed and physisorbed rates, after the last incident particle is launched onto the surface, we extend the calculations for up to 0.45~ns. The sticking coefficient for chemisorption is obtained based on the results of the $n=74$ incident particles hitting each of the $N=10$ amorphous carbon structures, as 
\begin{equation}
    P_\mathrm{x}=\frac{1}{N}\sum_{i=1}^{N}P_\mathrm{x}^{(i)}=\frac{1}{N}\sum_{i=1}^{N}\frac{N^{(i)}_\mathrm{x}}{n},
\end{equation}
where $P_\mathrm{x}^{(i)}$ is the sticking coefficient of the $i$-th structure for chemisorption (x$\equiv$chem) or physisorption (x$\equiv$phys), and $N^{(i)}_\mathrm{x}$ is the number of particles sticking to its surface through the process of interest, x. So, the sticking coefficient $P_\mathrm{x}$ is the mean of the distribution of $P_\mathrm{x}^{(i)}$. The error is computed as the variance of the mean sticking coefficients, obtained as
\begin{equation}
    \sigma_{P_\mathrm{x}}^2=\frac{1}{N}\sigma_{P_\mathrm{x}^{(i)}}^2=\frac{1}{N}\left(\frac{1}{N}\sum_{i=1}^N(P_\mathrm{x}^{(i)}-P_\mathrm{x})^2\right),
    \label{eq:sigma}
\end{equation}
with $\sigma_{P_\mathrm{x}^{(i)}}$ the variance of the distribution of $P_\mathrm{x}^{(i)}$. The results presented in section \ref{res_disc} are of the form $P_\mathrm{x}\pm\sigma_{P_\mathrm{x}}$.

\subsection{Binding energy estimations}

\begin{figure*}
\sidecaption
\includegraphics[width=12cm]{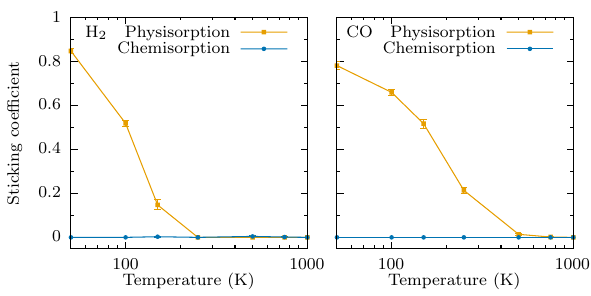} 
\caption{\textbf{Sticking coefficients as a function of temperature.} Temperature ranging from 50 to 1000~K, for H$_2$ (left panel) and CO (right panel) on an amorphous carbon structure.}
\label{fig:H2-CO}
\end{figure*}

We are interested in estimating the binding energies of the C, O, and H atoms on the amorphous carbon structure. To do this we use the energies calculated during the dynamics obtained through the simulations using ReaxFF (hence obtained with a force field). We first obtain the binding energy of C atoms, $E^\mathrm{C}_d$, by considering the energy of a whole amorphous carbon structure thermalized at any temperature (the binding energy is temperature-independent), divided by the number of C atoms in the structure, $N^\mathrm{C}$. Note that during the thermalization of the structure, some C atoms might desorb and hence $N^\mathrm{C}\leq 512$. For the other species X (X$\equiv$C or X$\equiv$H), we consider the total energy of the structure including the chemisorbed X atoms at the end of a simulation, when the structure is fully thermalized. We remove from this the energy coming from the C atoms and divide by the number of X atoms chemisorbed on the surface considered. 
\begin{equation}
    E^\mathrm{C}_d=\frac{E^\mathrm{tot}_d}{N^\mathrm{C}};\ \ \ \ \ E^\mathrm{X}_d=\frac{(E^\mathrm{tot}_d-N^\mathrm{C}E^\mathrm{C}_d)}{N^\mathrm{X}}.
\end{equation}
To obtain results independent of the choice of the structure and increase their accuracy, we reproduce this calculation for 9 different amorphous carbon structures, at 9 different temperatures. The error is computed as the variance of the mean value, similarly to Eqn.~\ref{eq:sigma}.

\section{Results and discussion}
\label{res_disc}
Here we first present our calculated sticking coefficients, and then binding energies that we obtain in order to estimate a desorption rate. We end this section by applying these sticking coefficients and binding energies to estimate dust growth rates in relevant astrophysical environments.
\subsection{Sticking coefficients}

We present the sticking coefficients of various colliders onto amorphous carbon structures. The results include physisorption, where the small Van der Waals interactions allow the collider to move around on the amorphous carbon structure without creating a chemical bond, and chemisorption, where the creation of a valence bond between the collider and the amorphous carbon structure allows the latter to grow by accretion. The physisorbed atoms, which are only present below about 250~K, can eventually desorb from the surface, on a much longer timescale at low temperature than at high temperature, with timescales of the order of nanoseconds down to a few picoseconds. The results belong to either of two groups: the non-reactive or the reactive colliders. 

\begin{figure}
\centering
\includegraphics[width=\columnwidth]{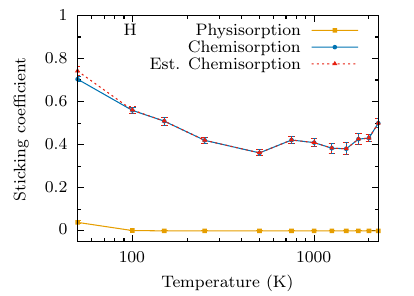} 
\caption{\textbf{Sticking coefficients as a function of temperature.} Temperature ranging from 50 to 1250~K for H on an amorphous carbon structure.}
\label{fig:H}
\end{figure}

\begin{figure}
\centering
\includegraphics[width=\columnwidth]{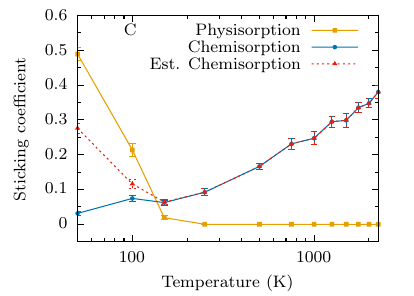} 
\caption{\textbf{Sticking coefficients as a function of temperature.} Temperature ranging from 50 to 1250~K for C on an amorphous carbon structure.}
\label{fig:C}
\end{figure}

In the non-reactive group, no chemisorption is observed, and the colliders are only physisorbed. These colliders hence cannot participate in the growth of cosmic dust. We found that among the colliders we consider in this study, H$_2$ and CO, presented respectively in the left and right panels of Fig.~\ref{fig:H2-CO}, are non-reactive colliders, as they cannot easily create covalent bonds with the surface. We do not observe any breaking of H$_2$, as the dissociation energy of its bond is above 400~kJ.mol$^{-1}$ (48110~K), or of CO, which was expected since it is the strongest known diatomic bond, with a dissociation energy of over 1000~kJ.mol$^{-1}$ (120270~K). For those two colliders, we present the sticking coefficients for physisorption over temperatures from 50~K to 1000~K. It is important to note that these coefficients are time-dependent, as the physisorbed H$_2$ and CO molecules will desorb after a given temperature-dependent time (longer for lower temperatures). Physisorption is restricted to temperatures of up to 600~K, as only Van der Waals forces bind the physisorbed species to the surface.

When H, C, or O is the collider, there is a possible sticking through chemisorption. These colliders hence actively participate in the growth of carbonaceous cosmic dust through collision followed by accretion. For those colliders, both chemisorption and physisorption can happen. For dust growth, only chemisorption will have significance. Hence, for low temperatures, where the coefficient for sticking by physisorption is very high, we estimate a temperature-dependent rate for the conversion from physisorption to either desorption or chemisorption, which yields an estimated long-time chemisorption. 

Fig.~\ref{fig:H} presents the temperature-dependent sticking coefficient of H on amorphous carbon. This coefficient is very high, particularly at low temperature, as it almost reaches 0.75 at 50~K. As the temperature increases, it quickly decreases to plateau around 0.4, and starts increasing again over 1500~K. It is interesting to note that over the 750-1500~K temperature range, the sticking coefficient appears to be temperature-independent. There is almost no physisorption even at 50~K, as the atomic hydrogen creates valence bonds very efficiently and, thus, rapidly chemisorbs.

\begin{figure}
\centering
\includegraphics[width=\columnwidth]{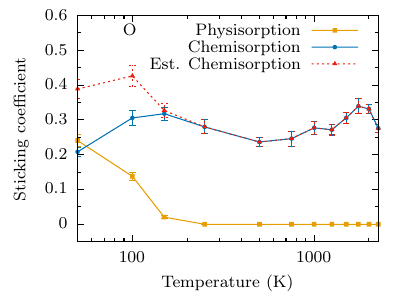} 
\caption{\textbf{Sticking coefficients as a function of temperature.} Temperature ranging from 50 to 1250~K for O on an amorphous carbon structure.}
\label{fig:O}
\end{figure}

The C and O sticking coefficients on the amorphous carbon are presented in Fig.~\ref{fig:C} and Fig.~\ref{fig:O}, respectively. The profiles of the H, C, and O sticking coefficients over temperature are very different qualitatively and quantitatively, emphasizing the importance of direct dynamical calculations compared to rough estimates. Considering only the initial chemisorbing fraction of C, it is clearly increasing with the temperature over the whole range considered, going from an almost non-reactive behavior at 50~K to a sticking coefficient of about 0.38 at 2250~K. On the other hand, the chemisorption rate of O is almost temperature independent, oscillating between 0.25 and 0.35 over the whole 50~K to 2250~K temperature range, with a slight decrease below 100~K. At low temperature (from 50~K to 250~K), we observe both chemisorption and physisorption. 

\begin{figure*}
\sidecaption
\includegraphics[width=12cm]{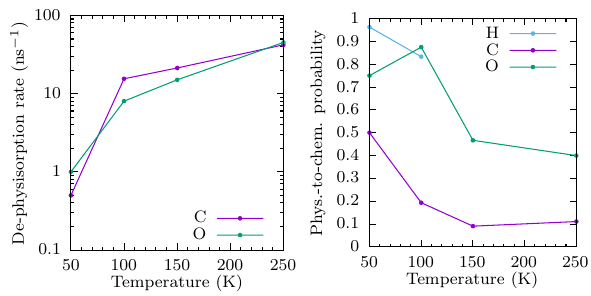} 
\caption{\textbf{Temperature-dependent quantities related to accretion.} Left panel: Temperature-dependent de-physisorption rate of the C and O atoms from the amorphous carbon structure. Right panel: Probability  that physisorbed H, C, and O atoms chemisorb onto the amorphous carbon structure as a function of temperature.}
\label{fig:lt-PtC}
\end{figure*}

We want to estimate a sticking coefficient including only chemisorption and no physisorption, considering that cosmic dust is in media diffuse enough that all physisorbed atoms will eventually chemisorb or desorb before being directly involved in another collision with a collider. This is confirmed in the left panel of Fig.~\ref{fig:lt-PtC}, which presents an estimated desorption rate (of physisorbed atoms) for C and O atoms. It is clear that physisorbed particles will not play a role in the growth of carbon dust because they remain physisorbed on a much shorter timescale ($\ll1$~s$^{-1}$) than the average collision time. 

We investigate how many of the physisorbed colliders will eventually desorb and how many will chemisorb. The ambient conditions do not change over the time scale during which this occurs. It happens in less than a few nanoseconds; hence, all parameters remain unchanged. We find it a good approximation to observe what happens during 0.45~ns and compute coefficients based on that to estimate the long-time result. In the right panel of Fig.~\ref{fig:lt-PtC} we present the probability that initially physisorbed atoms chemisorb onto the surface, $P_\mathrm{PtC}$. This probability is much higher for O atoms than for C atoms, and decreases with temperature, remaining relatively constant over 150~K. Note that the 50~K data point for O atoms is based on very low statistics and is therefore not as reliable as the other data presented. Similarly, there were not enough physisorbed H atoms to compute the probability over 100~K.

Based on $P_\mathrm{PtC}$, and on the sticking coefficient for chemisorption, $P_\mathrm{chem}$, and for physisorption, $P_\mathrm{phys}$, we can estimate the final sticking probability of chemisorption, $P_\mathrm{chem}^\mathrm{est}$, for H, C, and O atoms on amorphous carbon, where all atoms are chemisorbed or desorbed, without the need to perform numerically expensive calculations to converge those coefficients until the point where all physisorbed atoms desorb or chemisorb. The estimated chemisorption sticking coefficients (represented with dashed red lines in Fig.~\ref{fig:H}, \ref{fig:C} and \ref{fig:O}) are obtained as:
\begin{equation}
    P_\mathrm{chem}^\mathrm{est} = P_\mathrm{chem}+P_\mathrm{phys}\times P_\mathrm{PtC}
\end{equation}
The results at low temperature are drastically changed when the long-term behavior -- i.e., when all physisorbed atoms have either been chemisorbed or desorbed -- is considered. The physisorbed O atoms are efficiently chemisorbed, given enough time, and many C atoms are first physisorbed at low temperature, even if those are less efficiently chemisorbed afterward. This leads the estimated sticking coefficient for the chemisorption of C at 50~K to jump from 0.03 to 0.28, while that of O jumps from 0.2 to almost 0.4. The estimated chemisorption sticking coefficient is higher for O atoms than for C atoms on the 50~K-to-1000~K temperature range. Those coefficients are about equal between 1000~K and 2000~K, and the sticking of C atoms is more efficient than the sticking of O atoms above 2000~K, where the coefficient for the O atoms starts decreasing. This is an important result, as it shows that once the carbonaceous dust seed is created, O atoms will stick to it for most temperatures more likely than C atoms, hence changing the pure amorphous carbon structures over time to structures that include both C and O atoms in an environment where they are both present. However, carbonaceous dust forms mostly in the ejecta of C-rich stars rather than O-rich stars, so this effect will be reduced in these conditions because most O atoms will be trapped in CO molecules, which are non-reactive on carbonaceous dust grains. It is interesting to note that the sticking coefficient of O atoms is nearly temperature-independent over 150~K, while this is not the case for C atoms, for which sticking values steadily increase with the temperature over 150~K.

To make the results presented in this work accessible and usable, they are all given in Tab.~\ref{tab:Sc_all} of the appendix, along with their uncertainties.

\subsection{Binding energies}

In order to study the growth rate of amorphous carbon dust, we need to account for the thermal desorption of initially chemisorbed atoms, which depends on the temperature and the binding energy of the desorbing species. In this work, we estimate the binding energies that we need using a force field, as detailed in the section \ref{num_det}. Binding energies are presented in Tab.~\ref{tab:BE}, where carbon and oxygen atoms have larger binding energies than hydrogen atoms due to their ability to not only form single bonds. The binding energy for the C atom is consistent with results found in the literature, where binding energies of C atoms on graphene were found to be around 7-8~eV~\citep{oli_2013}. It can be noted that those values correspond to average binding energies, while some atoms might have higher or lower binding energies, depending on the number of bonds formed with the dust grain and on the local structure of the grain around the binding region. It can be expected that the more weakly adsorbed species will desorb preferentially. The binding energies we computed here may therefore be a slight overestimate for the purposes of calculating thermal desorption rates. 

\begin{table}
   \centering
   \topcaption{\textbf{Binding energies.} Computed for H, C, and O with C atoms of the amorphous carbon structure.} 
   \begin{tabular}{@{} lccc @{}} 
      \toprule
      Species X & H & C & O \\
      \midrule\midrule
	   Binding energy $E^\mathrm{X}_d$ (eV) & 4.76$\pm$0.12 & 7.55$\pm$0.01 & 7.68$\pm$0.21 \\
      \bottomrule
   \end{tabular}
\label{tab:BE}
\end{table}

\subsection{Application to dust growth}

In stellar ejecta where dust grains form, molecules and dust form through nucleation and condensation of gas-phase atoms \citep{sarangi2018book}. Amorphous carbon or graphite dust is known to form in regions where the C/O ratio is typically greater than 1 \citep{fabbian_2009}, for example in the He-shell of core-collapse supernovae \citep{sarangi2018book,sarangi_2022b} or the winds of C-rich AGB stars \citep{nanni_2019,marini_2021}. In such environments we expect to find atoms and molecules such as C, O, H, CO, He, H$_2$, CO$_2$, CH etc. In this paper, we explore the probability that several of these species -- namely H, H$_2$, C, O and CO -- stick to the surface of a given dust grain. 

For a grain of radius $a$, the rate at which it will accrete a particular gas-phase species on its surface depends on (a) its geometric cross-section, (b) the velocity of the incoming gas-phase species (which depends on the gas temperature $T_\mathrm{g}$, (c) the density of that species in the gas, and also (d) the efficiency of sticking on the surface, which is commonly called the sticking coefficient. This accretion coefficient is defined as the rate of sticking $F_i$ of a given species $i$ (see Eqn.~\ref{eq_rate_acc}), as a function of the dust grain radius $a$ and temperature $T_\mathrm{d}$ and the density of species $n_i$ \citep{dwe11, hirashita_2012}.
\begin{equation}
\label{eq_rate_acc}
   F_i(a,n_i,T_\mathrm{d},T_\mathrm{g},t) = \pi a^2 S_i(T_\mathrm{d},T_\mathrm{g}) n_i(t) v^{th}_i(T_\mathrm{g})
\end{equation}
where $S_i(T_\mathrm{d},T_\mathrm{g})$ is the temperature-dependent sticking coefficient of gas species at temperature $T_\mathrm{g}$ on a dust grain at temperature $T_\mathrm{d}$, and $v^{th}_i(T_\mathrm{g})$ is the thermal velocity of the species $i$, here at the same temperature $T_\mathrm{d}=T_\mathrm{g}=T$ as the dust grain. Combining the accretion rate of all species in the gas-phase, the rate at which the mass of a dust grain $m(a,t)$ will change is therefore given by
\begin{equation}
   \frac{\mathrm{d}m (a,t)}{\mathrm{d}t} = \sum_i  m_i F_i
\end{equation}
with $m_i$ the mass of species $i$. When atoms or molecules accrete on the surface of the dust grains, their abundances in the gas drop at the same rate \citep{hirashita_2011}. If we assume a general time-dependence of the gas density (without accretion) in a given environment as $n \sim t^p$, then after accretion on the dust-surface, the remaining species of type $i$ in the gas is given by,
\begin{equation}
\label{eq_n_i}
n_i(t+\Delta t) = (n_i(t)- n_d(t) F_i\Delta t) \Big(1 + \frac{\Delta t}{t}\Big)^p
\end{equation}
where $n_d(t)$ is the number density of dust grains in the gas, since each of those grains will accrete at the same rate. In any freely expanding gas, such as stellar winds or explosions, the power law exponent $p$ is given by $-3$.  

\begin{figure*}[htbp]
\centering
\includegraphics[width=7.6cm]{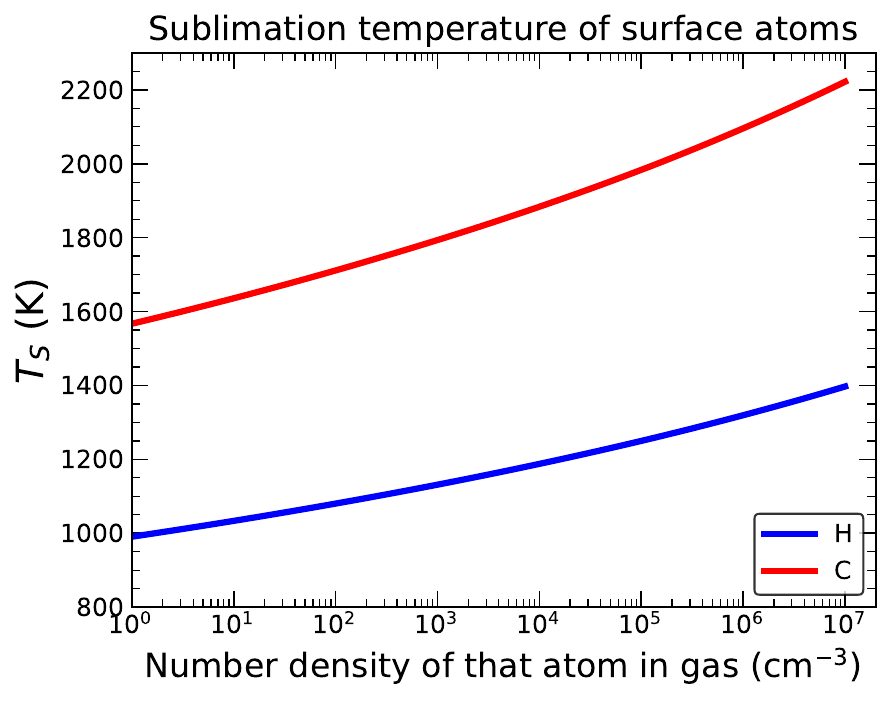} 
\includegraphics[width=7.6cm]{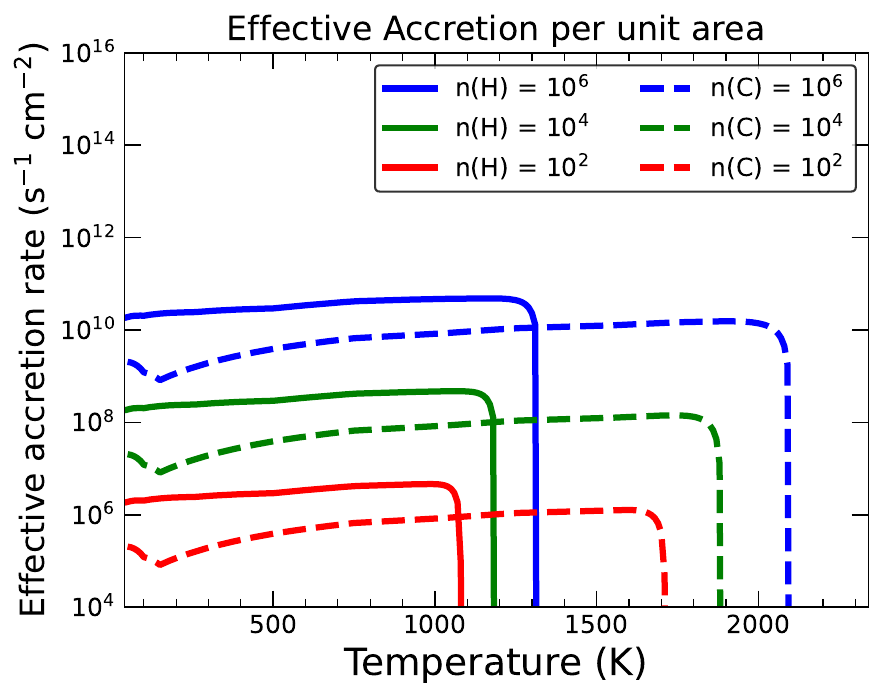} 
\caption{\textbf{Carbonaceous dust growth.} Left panel: Sublimation temperature, where the accretion and desorption rates on the surface of amorphous carbonaceous dust grains balance each other out, shown as a function of densities of C and H atoms in the gas. Right panel: Effective accretion rates as a function of temperature on the surface of amorphous carbonaceous dust grains (effective number of accreting atoms per unit time per unit area), shown for different densities of C and H atoms in the gas. This is therefore the grain growth rate for the given conditions. }
\label{fig:dustgrowth}
\end{figure*}

In parallel to accretion, the dust grains also undergo desorption of atoms from their surface because the atoms may leave the grains when they gain enough energy to overcome the binding energy. The rate $D_i$ for the desorption of species $i$ strongly depends on the dust temperature and can be expressed by the Polanyi–Wigner equation \citep{potapov_2018}, as
\begin{equation}
\label{eq_desorption}
D_i(a,T_\mathrm{d},t) = \Lambda_i(a) \ \nu_i \ e^{(-E^i_{d}/k_BT_\mathrm{d})}
\end{equation}
where, $\Lambda_i$ is the number of surface atoms, $\nu_i$ is the attempted frequency for escape, $E^i_d$ is the binding energy of species $i$, and $k_B$ is the Boltzmann constant. Comparing the rate of accretion to that of desorption, we find the temperature where these two rates are equal, 
\begin{equation}
    F_i(a,n_i,T_S,T_\mathrm{g},t) = D_i(a,T_S)
\end{equation}
which represents the sublimation temperature $T_S$. In the conditions considered here, the temperature of the gas and dust is the same and hence $T_\mathrm{g}=T_S$. Above this temperature, the grains rapidly loose atoms and sublimate.

In circumstellar environments, where carbonaceous dust can form, C, H, He, O, CO, and H$_2$ are often abundant. When the C/O ratio is larger than 1, there are almost no O atoms left in the gas by the time the smallest grains of carbonaceous dust are formed. The O atoms are trapped in CO molecules \citep{cherchneff_2000a, cherchneff_2000b, sarangi2018book}. As Fig.~\ref{fig:H2-CO} clearly indicates, the efficiency of CO and H$_2$ sticking on a surface of amorphous carbonaceous dust is almost negligible. The efficiencies of sticking of H and C atoms are shown in Figs.~\ref{fig:H} and \ref{fig:C} respectively.

For the frequency $\nu_i$ in the desorption rate (see Eqn. \ref{eq_desorption}), we use the vibrational frequency of the C-C and C-H bonds as 3$\times$10$^{13}$ and 9$\times$10$^{13}$~s$^{-1}$ respectively \citep{stuart_2004}. Since both accretion and desorption rates have the same dependency on the surface area of the dust grains, the sublimation temperature $T_S$ is therefore independent of the grain size. In Fig.~\ref{fig:dustgrowth} (left panel), the sublimation temperature is shown for the C and H atoms on our amorphous carbonaceous dust grains as a function of their respective density in the gas-phase. The C atom has a larger sublimation temperature given its larger binding energy. 

As the density of atoms drop in the gas, the sublimation temperature declines, since there are less atoms in the gas to accrete on the surface that can balance the desorption process. We call the difference between the accretion and the desorption rate the effective accretion rate. 
Fig.~\ref{fig:dustgrowth} (right panel) presents the effective accretion rate of C and H atoms for different densities, based on the individual sticking coefficients calculated in this study. As shown above and consistent with the figure, the efficiency of H-sticking on our amorphous carbonaceous dust grains is greater than that of C-sticking, the difference being more pronounced at lower temperatures. More importantly, at higher temperatures, the H atoms desorb much faster than the C atoms.
Therefore, in a cooling gas, like most stellar winds or ejecta, the initial phase of the grain growth will be dominated by dust made purely from C atoms, and C atoms will get depleted in the gas through sticking on the surface of the grains. As the gas further cools, the H atoms, based on how abundant H is, will start accreting on the surface of the grains, rapidly forming new hydrocarbons and possibly polyaromatic hydrocarbons (PAHs). It is important to note that a fraction of H is also bound in molecular H$_2$, which has a very low sticking probability. The temperature evolution of the environment has a balancing impact on the effective accretion rates. If the accretion rate is high, the atoms are rapidly depleted from the gas, which then leads to a lower sublimation temperature and more desorption, until the temperature further cools down.

\section{Conclusion}
\label{concl}

This study is unique because: (a) we have, for the first time, explored the variation in the efficiency of sticking for different atoms and molecules on an amorphous carbonaceous dust grain; (b) we have studied the dynamics of grain accretion from the gas-phase over a wide temperature range, applicable to astrophysical environments in which carbonaceous dust nucleation is expected to occur, namely the ejecta of supernovae and AGB stars (c) we have developed the necessary tools to quantify the rate of grain growth through accretion of gas-phase species on a dust surface. This study is particularly important where dust seeds are synthesized from gas-phase clusters, and molecule and dust formation occur simultaneously. The abundances of the species which can stick on the surface therefore can be complemented by chemical reactions and sticking rates. We choose here to limit the accuracy of the sticking coefficients to $1.10^{-3}$, as the calculations would be too computationally expensive to reach a higher accuracy (10 times more collisions are necessary to compute sticking coefficients with an accuracy of $1.10^{-4}$). This therefore limits this study to the most abundant species found in the media of interest, as the gas species with relative abundances to the main species of $10^{-4}$ or lower will have a contribution at the limit of the sensitivity of our model. This is not a problem as the most abundant species will be the ones that contribute to the growth of carbonaceous dust.

We found that atoms such as H, O, and C have sticking probabilities on a carbonaceous dust grain that range between 0.1 and 0.75. In contrast, H$_2$ and CO molecules have an almost negligible probability of sticking to dust grains. Dust formation in stellar winds occurs concurrently with gas-phase nucleation processes \citep{cherchneff_iau1999, sarangi2018book}. The specific abundances of carbon, oxygen, and hydrogen vary from star to star and between supernovae. In addition, the physical conditions, mainly the gas temperature and densities, define the relative efficiency between routes of dust production and molecular formation. In environments where carbon dust forms, most of the O atoms are locked up in CO molecules. Moreover, the gas temperature at which CO molecules may form is much higher than the sublimation temperature of carbonaceous dust~\citep{sarangi2018book}. Hence, in a cooling gas, before the smallest carbon dust grains are formed, all the O atoms are depleted. In this study, we show that, even though O atoms have a high sticking efficiency, the sticking efficiency of CO molecules is nearly zero. So amorphous carbon dust grains are not expected to accrete O atoms on their surfaces in these environments, neither in atomic nor in molecular form.

Grain growth through accretion of C and H atoms depends on the number of available atoms in the gas, after the smallest dust grains or dust-seeds are formed. Chemical modeling \citep{sarangi2018book} suggests that C atoms often tend to form small clusters of C-chains in the gas-phase. Based on the present results on the sticking of H$_2$ and CO, we find it unlikely that small molecules with saturated valences are efficient in aiding the grain growth process via accretion on the dust surface.

Our study points out that the difference in effective accretion rates of C and H may shape the structure of a circumstellar or interstellar grain of carbon dust. The strong temperature dependence of H-sticking and C-sticking reported in this study, in the future, could help us understand the aromaticity of carbon-dust in the ISM and the routes for PAH condensation \citep{dartois_2007}. 

This study will be further extended to silicate-type dust, that is found nearby O-rich stars, for which the dust grain structure is more challenging, and includes various chemical species. It will also be of great interest to the astrophysical modeling community to provide sticking coefficients for conditions where the temperature of the gas and of the dust differ, for both carbonaceous and silicate grains. This is particularly needed in the interstellar medium and gas clouds, where the dust temperature remains low, while incident species can have temperatures spanning a large range. This will be considered in future work, where it will be interesting to see which temperature -- dust or gas -- has a greater influence on the sticking process.

\bibliography{bibtex/Dust.bib} 

\begin{thebibliography}{48}
\expandafter\ifx\csname natexlab\endcsname\relax\def\natexlab#1{#1}\fi

\bibitem[{Al-Halabi {et~al.}(2004)Al-Halabi, van Dishoeck, \&
  Kroes}]{alhalabi_jcp2004}
Al-Halabi, A., van Dishoeck, E.~F., \& Kroes, G.~J. 2004, J. Chem. Phys., 120,
  3358

\bibitem[{Andersen {et~al.}(2003)Andersen, H{\"o}fner, \&
  Gautschy-Loidl}]{andersen_iau2003}
Andersen, A., H{\"o}fner, S., \& Gautschy-Loidl, R. 2003, Proc. Int. Astron.
  Union, 210, A13

\bibitem[{{Bocchio, Marco} {et~al.}(2014){Bocchio, Marco}, {Jones, Anthony P.},
  \& {Slavin, Jonathan D.}}]{bocchio_aa2014}
{Bocchio, Marco}, {Jones, Anthony P.}, \& {Slavin, Jonathan D.} 2014, A\&A,
  570, A32

\bibitem[{{Buch} \& {Zhang}(1991)}]{buch_apj1991}
{Buch}, V. \& {Zhang}, Q. 1991, ApJ, 379, 647

\bibitem[{{Cazaux, S.} {et~al.}(2011){Cazaux, S.}, {Morisset, S.}, {Spaans,
  M.}, \& {Allouche, A.}}]{cazaux_aa2011}
{Cazaux, S.}, {Morisset, S.}, {Spaans, M.}, \& {Allouche, A.} 2011, A\&A, 535,
  A27

\bibitem[{{Chaabouni, H.} {et~al.}(2012){Chaabouni, H.}, {Bergeron, H.},
  {Baouche, S.}, {Dulieu, F.}, {Matar, E.}, {Congiu, E.}, {Gavilan, L.}, \&
  {Lemaire, J. L.}}]{chaabouni_aa2012}
{Chaabouni, H.}, {Bergeron, H.}, {Baouche, S.}, {et~al.} 2012, A\&A, 538, A128

\bibitem[{Chenoweth {et~al.}(2008)Chenoweth, van Duin, \&
  Goddard}]{chenoweth_jpca2008}
Chenoweth, K., van Duin, A. C.~T., \& Goddard, W.~A. 2008, J. Phys. Chem. A,
  112, 1040

\bibitem[{{Cherchneff}(2000)}]{cherchneff_2000a}
{Cherchneff}, I. 2000, in The Carbon Star Phenomenon, ed. R.~F. {Wing}, Vol.
  177, 331

\bibitem[{Cherchneff \& Cau(1999)}]{cherchneff_iau1999}
Cherchneff, I. \& Cau, P. 1999, Proc. Int. Astron. Union, 191, 251–260

\bibitem[{{Cherchneff} {et~al.}(2000){Cherchneff}, {Le Teuff}, {Williams}, \&
  {Tielens}}]{cherchneff_2000b}
{Cherchneff}, I., {Le Teuff}, Y.~H., {Williams}, P.~M., \& {Tielens},
  A.~G.~G.~M. 2000, A\&A, 357, 572

\bibitem[{{Dartois} {et~al.}(2007){Dartois}, {Geballe}, {Pino}, {Cao}, {Jones},
  {Deboffle}, {Guerrini}, {Br{\'e}chignac}, \& {D'Hendecourt}}]{dartois_2007}
{Dartois}, E., {Geballe}, T.~R., {Pino}, T., {et~al.} 2007, A\&A, 463, 635

\bibitem[{de~Rooij {et~al.}(2010)de~Rooij, Kleyn, \& Goedheer}]{rooij_pccp2010}
de~Rooij, E.~D., Kleyn, A.~W., \& Goedheer, W.~J. 2010, Phys. Chem. Chem.
  Phys., 12, 14067

\bibitem[{Deringer {et~al.}(2018)Deringer, Caro, Jana, Aarva, Elliott, Laurila,
  Csányi, \& Pastewka}]{deringer_CM2018}
Deringer, V.~L., Caro, M.~A., Jana, R., {et~al.} 2018, Chem. Mater., 30, 7438

\bibitem[{Deringer \& Cs\'anyi(2017)}]{deringer_PRB2017}
Deringer, V.~L. \& Cs\'anyi, G. 2017, Phys. Rev. B, 95, 094203

\bibitem[{{Draine}(2011)}]{draine_2011}
{Draine}, B.~T. 2011, {Physics of the Interstellar and Intergalactic Medium}
  (Princeton University Press)

\bibitem[{{Dwek} \& {Cherchneff}(2011)}]{dwe11}
{Dwek}, E. \& {Cherchneff}, I. 2011, ApJ, 727, 63

\bibitem[{{Dwek} \& {Scalo}(1980)}]{dwek_1980}
{Dwek}, E. \& {Scalo}, J.~M. 1980, ApJ, 239, 193

\bibitem[{Ehrenfreund \& Cami(2010)}]{ehrenfreund_cshlp2010}
Ehrenfreund, P. \& Cami, J. 2010, Cold Spring Harb. perspect. biol., 2

\bibitem[{{Fabbian} {et~al.}(2009){Fabbian}, {Nissen}, {Asplund}, {Pettini}, \&
  {Akerman}}]{fabbian_2009}
{Fabbian}, D., {Nissen}, P.~E., {Asplund}, M., {Pettini}, M., \& {Akerman}, C.
  2009, A\&A, 500, 1143

\bibitem[{Gobrecht {et~al.}(2019)Gobrecht, Plane, Bromley, Decin, \&
  Cristallo}]{gobrecht_iau2019}
Gobrecht, D., Plane, J.~M., Bromley, S.~T., Decin, L., \& Cristallo, S. 2019,
  Proc. Int. Astron. Union, 15, 245–248

\bibitem[{Henning(2010)}]{henning_araa2010}
Henning, T. 2010, ARA\&A, 48, 21

\bibitem[{{Hirashita}(2012)}]{hirashita_2012}
{Hirashita}, H. 2012, MNRAS, 422, 1263

\bibitem[{{Hirashita} \& {Kuo}(2011)}]{hirashita_2011}
{Hirashita}, H. \& {Kuo}, T.-M. 2011, MNRAS, 416, 1340

\bibitem[{Hollenbach \& Salpeter(1970)}]{hollenbach_jcp1970}
Hollenbach, D. \& Salpeter, E.~E. 1970, J. Chem. Phys., 53, 79

\bibitem[{{Hollenbach} \& {Salpeter}(1971)}]{hollenbach_apj1971}
{Hollenbach}, D. \& {Salpeter}, E.~E. 1971, ApJ, 163, 155

\bibitem[{Hoover(1985)}]{hoover_pra1985}
Hoover, W.~G. 1985, Phys. Rev. A, 31, 1695

\bibitem[{{Jones} \& {Nuth}(2011)}]{jones_2011}
{Jones}, A.~P. \& {Nuth}, J.~A. 2011, A\&A, 530, A44

\bibitem[{{Laffon} {et~al.}(2021){Laffon}, {Ferry}, {Grauby}, \&
  {Parent}}]{laffon_2021}
{Laffon}, C., {Ferry}, D., {Grauby}, O., \& {Parent}, P. 2021, Nat. Astron., 5,
  445

\bibitem[{{Leitch-Devlin} \& {Williams}(1985)}]{devlin_1985}
{Leitch-Devlin}, M.~A. \& {Williams}, D.~A. 1985, MNRAS, 213, 295

\bibitem[{Marchione {et~al.}(2019)Marchione, Rosu-Finsen, Taj, Lasne,
  Abdulgalil, Thrower, Frankland, Collings, \& McCoustra}]{marchione_aesc2019}
Marchione, D., Rosu-Finsen, A., Taj, S., {et~al.} 2019, ACS Earth Space Chem.,
  3, 1915

\bibitem[{{Marini} {et~al.}(2021){Marini}, {Dell'Agli}, {Groenewegen},
  {Garc{\'\i}a-Hern{\'a}ndez}, {Mattsson}, {Kamath}, {Ventura}, {D'Antona}, \&
  {Tailo}}]{marini_2021}
{Marini}, E., {Dell'Agli}, F., {Groenewegen}, M.~A.~T., {et~al.} 2021, A\&A,
  647, A69

\bibitem[{Masuda \& Takahashi(1997)}]{masuda_asr1997}
Masuda, K. \& Takahashi, J. 1997, Adv. Space Res., 19, 1019

\bibitem[{{Nanni} {et~al.}(2019){Nanni}, {Groenewegen}, {Aringer}, {Rubele},
  {Bressan}, {van Loon}, {Goldman}, \& {Boyer}}]{nanni_2019}
{Nanni}, A., {Groenewegen}, M. A.~T., {Aringer}, B., {et~al.} 2019, MNRAS, 487,
  502

\bibitem[{Nashimoto {et~al.}(2020)Nashimoto, Hattori, Poidevin, \&
  Génova-Santos}]{nashimoto_apjl2020}
Nashimoto, M., Hattori, M., Poidevin, F., \& Génova-Santos, R. 2020, ApJ, 900,
  L40

\bibitem[{Nosé(1984)}]{nose_jcp1984}
Nosé, S. 1984, J. Chem. Phys., 81, 511

\bibitem[{Oli {et~al.}(2013)Oli, Bhattarai, Nepal, \& Adhikari}]{oli_2013}
Oli, B.~D., Bhattarai, C., Nepal, B., \& Adhikari, N.~P. 2013, in Advanced
  Nanomaterials and Nanotechnology (Springer Berlin Heidelberg), 515--529

\bibitem[{{Potapov} {et~al.}(2018){Potapov}, {J{\"a}ger}, \&
  {Henning}}]{potapov_2018}
{Potapov}, A., {J{\"a}ger}, C., \& {Henning}, T. 2018, ApJ, 865, 58

\bibitem[{Potapov {et~al.}(2020)Potapov, J\"ager, \& Henning}]{potapov_prl2020}
Potapov, A., J\"ager, C., \& Henning, T. 2020, Phys. Rev. Lett., 124, 221103

\bibitem[{Potapov \& McCoustra(2021)}]{potapov_irpc2021}
Potapov, A. \& McCoustra, M. 2021, Int. Rev. Phys. Chem., 40, 299

\bibitem[{{Sarangi}(2022)}]{sarangi_2022b}
{Sarangi}, A. 2022, A\&A, 668, A57

\bibitem[{{Sarangi} {et~al.}(2018){Sarangi}, {Matsuura}, \&
  {Micelotta}}]{sarangi2018book}
{Sarangi}, A., {Matsuura}, M., \& {Micelotta}, E.~R. 2018, Space Sci. Rev.,
  214, 63

\bibitem[{{Stuart}(2004)}]{stuart_2004}
{Stuart}, B.~H. 2004, {Infrared Spectroscopy: Fundamentals and Applications}
  (John Wiley and Sons, Ltd)

\bibitem[{Tielens(2022)}]{tielens_fass2022}
Tielens, A. 2022, Front. Astron. Space Sci., 9

\bibitem[{{Tielens}(1998)}]{tielens_1998}
{Tielens}, A.~G.~G.~M. 1998, ApJ, 499, 267

\bibitem[{van Duin {et~al.}(2001)van Duin, Dasgupta, Lorant, \&
  Goddard}]{duin_jpca2001}
van Duin, A. C.~T., Dasgupta, S., Lorant, F., \& Goddard, W.~A. 2001, J. Phys.
  Chem. A, 105, 9396

\bibitem[{van Duin {et~al.}(2023)van Duin, Goddard, Islam, van Schoot, Trnka,
  \& Yakovlev}]{reaxFF_SCM}
van Duin, A. C.~T., Goddard, W.~A., Islam, M.~M., {et~al.} 2023, {ReaxFF
  2023.1, SCM, Theoretical Chemistry, Vrije Universiteit, Amsterdam, The
  Netherlands}

\bibitem[{Veeraghattam {et~al.}(2014)Veeraghattam, Manrodt, Lewis, \&
  Stancil}]{veeraghattam_aj2014}
Veeraghattam, V.~K., Manrodt, K., Lewis, S.~P., \& Stancil, P.~C. 2014, ApJ,
  790, 4

\bibitem[{{von Keudell} {et~al.}(2002){von Keudell}, Meier, \&
  Hopf}]{vonkeudell_2002}
{von Keudell}, A., Meier, M., \& Hopf, C. 2002, Diam. Relat. Mater., 11, 969

\end{thebibliography}
\bibliographystyle{bibtex/aa.bst} 

\begin{acknowledgements}
This work was supported by the Knut and Alice Wallenberg Foundation under grant nr. KAW 2020.0081.
The computations were enabled by resources provided by Chalmers e-Commons at Chalmers.
\end{acknowledgements}

\begin{appendix}
\section{Additional table}

\begin{sidewaystable*}
   \centering
   \topcaption{\centering\textbf{Sticking coefficients values.} Given for physisorption, chemisorption, and estimated chemisorption (asymptotic value) processes for H, H$_2$, C, O and CO \\colliding with amorphous carbon.\hspace{16.45cm}~} 
   \begin{tabular}{@{} lrcccccccccccr @{}} 
      \toprule
      \multicolumn{2}{c}{} & \multicolumn{12}{c}{Temperature (K)} \\
      \cmidrule(l){3-14} 
      Collider & Process & 50 & 100 & 150 & 250 & 500 & 750 & 1000 & 1250 & 1500 & 1750 & 2000 & 2250 \\
      \midrule\midrule
	H & Physis. & 0.039 & 0.001 & 0.000& 0.000& 0.000& 0.000& 0.000& 0.000& 0.000& 0.000& 0.000& 0.000\\
      &  & $\pm$0.007 & $\pm$0.001 & $\pm$0.000 & $\pm$0.000 & $\pm$0.000 & $\pm$0.000 & $\pm$0.000 & $\pm$0.000 & $\pm$0.000 & $\pm$0.000 & $\pm$0.000 & $\pm$0.000 \\
      \cmidrule(l){2-14}
      & Chemis. & 0.704 & 0.559 & 0.509 & 0.420 & 0.362 & 0.422 & 0.409 & 0.384 & 0.381 & 0.426 & 0.431 & 0.500 \\
      &  & $\pm$0.012 & $\pm$0.015 & $\pm$0.019 & $\pm$0.013 & $\pm$0.014 & $\pm$0.017 & $\pm$0.019 & $\pm$0.024 & $\pm$0.028 & $\pm$0.026 & $\pm$0.018 & $\pm$0.021 \\
      \cmidrule(l){2-14}
      & Est. Chemis. & 0.742 & 0.561 & 0.509 & 0.420 & 0.362 & 0.422 & 0.409 & 0.384 & 0.381 & 0.426 & 0.431 & 0.500 \\
      &  & $\pm$0.019 & $\pm$0.016 & $\pm$0.019 & $\pm$0.013 & $\pm$0.014 & $\pm$0.017 & $\pm$0.019 & $\pm$0.024 & $\pm$0.028 & $\pm$0.026 & $\pm$0.018 & $\pm$0.021 \\
      \midrule
    H$_2$ & Physis. & 0.846 & 0.518 & 0.150 & 0.000& 0.000& 0.000& 0.000&  &  &  &  & \\
      &  & $\pm$0.013 & $\pm$0.014 & $\pm$0.023 & $\pm$0.000 & $\pm$0.000 & $\pm$0.000 & $\pm$0.000 &  &  &  &  & \\
      \cmidrule(l){2-14}
      & Chemis. & 0.000& 0.000& 0.003 & 0.000& 0.004 & 0.001 & 0.000&  &  &  &  & \\
      &  & $\pm$0.000 & $\pm$0.000 & $\pm$0.002 & $\pm$0.000 & $\pm$0.002 & $\pm$0.001 & $\pm$0.000 &  &  &  &  & \\
      \midrule
    C & Physis. & 0.489 & 0.214 & 0.019 & 0.000& 0.000& 0.000& 0.000& 0.000& 0.000& 0.000& 0.000& 0.000\\
      &  & $\pm$0.018 & $\pm$0.018 & $\pm$0.006 & $\pm$0.000 & $\pm$0.000 & $\pm$0.000 & $\pm$0.000 & $\pm$0.000 & $\pm$0.000 & $\pm$0.000 & $\pm$0.000 & $\pm$0.000 \\
      \cmidrule(l){2-14}
      & Chemis. & 0.031 & 0.074 & 0.062 & 0.092 & 0.166 & 0.231 & 0.247 & 0.295 & 0.299 & 0.335 & 0.347 & 0.380 \\
      &  & $\pm$0.005 & $\pm$0.009 & $\pm$0.008 & $\pm$0.010 & $\pm$0.008 & $\pm$0.016 & $\pm$0.018 & $\pm$0.016 & $\pm$0.020 & $\pm$0.015 & $\pm$0.013 & $\pm$0.018 \\
      \cmidrule(l){2-14}
      & Est. Chemis. & 0.276 & 0.116 & 0.064 & 0.092 & 0.166 & 0.231 & 0.247 & 0.295 & 0.299 & 0.335 & 0.347 & 0.380 \\
      &  & $\pm$0.005 & $\pm$0.013 & $\pm$0.008 & $\pm$0.010 & $\pm$0.008 & $\pm$0.016 & $\pm$0.018 & $\pm$0.016 & $\pm$0.020 & $\pm$0.015 & $\pm$0.013 & $\pm$0.018 \\
      \midrule
    O & Physis. & 0.241 & 0.138 & 0.020 & 0.000& 0.000& 0.000& 0.000& 0.000& 0.000& 0.000& 0.000& 0.000\\
      &  & $\pm$0.018 & $\pm$0.012 & $\pm$0.005 & $\pm$0.000 & $\pm$0.000 & $\pm$0.000 & $\pm$0.000 & $\pm$0.000 & $\pm$0.000 & $\pm$0.000 & $\pm$0.000 & $\pm$0.000 \\
      \cmidrule(l){2-14}
      & Chemis. & 0.208 & 0.305 & 0.318 & 0.280 & 0.236 & 0.246 & 0.277 & 0.272 & 0.305 & 0.339 & 0.331 & 0.276 \\
      &  & $\pm$0.013 & $\pm$0.020 & $\pm$0.018 & $\pm$0.020 & $\pm$0.013 & $\pm$0.021 & $\pm$0.018 & $\pm$0.015 & $\pm$0.017 & $\pm$0.022 & $\pm$0.012 & $\pm$0.011 \\
      \cmidrule(l){2-14}
      & Est. Chemis. & 0.389 & 0.426 & 0.327 & 0.280 & 0.236 & 0.246 & 0.277 & 0.272 & 0.305 & 0.339 & 0.331 & 0.276 \\
      &  & $\pm$0.027 & $\pm$0.031 & $\pm$0.021 & $\pm$0.020 & $\pm$0.013 & $\pm$0.021 & $\pm$0.018 & $\pm$0.015 & $\pm$0.017 & $\pm$0.022 & $\pm$0.012 & $\pm$0.011 \\
      \midrule
    CO & Physis. & 0.780 & 0.658 & 0.516 & 0.214 & 0.014 & 0.001 & 0.000&  &  &  &  & \\
      &  & $\pm$0.019 & $\pm$0.014 & $\pm$0.021 & $\pm$0.013 & $\pm$0.004 & $\pm$0.001 &  &  &  &  &  & \\
      \cmidrule(l){2-14}
      & Chemis. & 0.000& 0.000& 0.000& 0.000& 0.000& 0.000& 0.000&  &  &  &  & \\
      &  & $\pm$0.000 & $\pm$0.000 & $\pm$0.000 & $\pm$0.000 & $\pm$0.000 & $\pm$0.000 & $\pm$0.000 &  &  &  &  & \\
      \bottomrule
   \end{tabular}
\label{tab:Sc_all}
\end{sidewaystable*}

\end{appendix}

\label{LastPage}
\end{document}